**Aristides Moustakas[1,*] and Matthew R. Evans[1]**

1. School of Biological and Chemical Sciences
Queen Mary University of London
Mile End Road, London, E1 4NS, UK

* Corresponding author

## Abstract

Bovine TB is a major problem for the agricultural industry in several countries. TB can be contracted and spread by species other than cattle and this can cause a problem for disease control. In the UK and Ireland, badgers are a recognised reservoir of infection and there has been substantial discussion about potential control strategies. We present a coupling of individual based models of bovine TB in badgers and cattle, which aims to capture the key details of the natural history of the disease and of both species at approximately county scale. The model is spatially explicit it follows a very large number of cattle and badgers on a different grid size for each species and includes also winter housing. We show that the model can replicate the reported dynamics of both cattle and badger populations as well as the increasing prevalence of the disease in cattle. Parameter space used as input in simulations was swept out using Latin hypercube sampling and sensitivity analysis to model outputs was conducted using mixed effect models. By exploring a large and computationally intensive parameter space we show that of the available control strategies it is the frequency of TB testing and whether or not winter housing is practised that have the most significant effects on the number of infected cattle, with the effect of winter housing becoming stronger as farm size increases. Whether badgers were culled or not explained about 5%, while the accuracy of the test employed to detect infected cattle explained less than 3% of the variance in the number of infected cattle.

## Keywords

## Introduction

Historically, *bovine tuberculosis* (TB, also abbreviated as bTB) was once almost eradicated from Great Britain, with a minimum in the number of cattle herds that contained an individual that reacted positively to a test for TB being reached in the late 1970s (Krebs et al., 1997). Since this time the disease has been steadily increasing in both its prevalence in the cattle herd and its spread around the country (Gilbert et al., 2005). Concern about the increasing prevalence of TB led to the Krebs Review, which concluded that there was compelling evidence for badgers acting as a wild reservoir host and implicating them in the transmission of TB to cattle (Krebs et al., 1997).

The Krebs Review proposed a large-scale field trial, the Randomised Badger Culling Trial (RBCT), to quantify the impact of culling badgers on the incidence of TB in cattle and the effectiveness of alternative culling strategies. The RBCT was implemented and took place between 1998 and 2007. The results of this experimental trial were analysed and published by the Independent Scientific Group (ISG) (Bourne et al., 2007; Godfray et al., 2004). The ISG concluded that badgers contribute significantly to increasing the incidence of the disease in cattle and that there is a cycle of infection between cattle and badgers. The ISG suggested that it would be difficult to eradicate TB completely as the test employed to detect TB in cattle produced false negative results (i.e. some individuals test negative despite being infected with TB) (Claridge et al., 2012; Szmaragd et al., 2012), and disease was re-introduced by badgers (i.e. badgers acted as a reservoir of infection). The ISG and subsequent analyses have suggested that proactive culling of badgers (killing badgers within an area whether or not there were any recorded outbreaks of TB) would be highly likely to reduce the incidence of TB within and around the culled area as long as the culled area was at least 141km$^2$ (Godfray et al., 2004; Jenkins et al., 2010). However, the estimated costs of culling at this scale have been regarded as too large to justify the likely economic benefits (Wilkinson et al., 2009). In addition, badgers within culling areas may show post-cull perturbation resulting in higher motility that risks spreading the disease further than in the un-culled state (Woodroffe et al., 2006a). In 2007, the then Chief Scientific Advisor for the UK, produced a report focussing on the scientific issues relating to the role that badger culling could play in controlling and reducing the levels of cattle TB. The conclusion of this report was that the removal of badgers was the best option available at that time to reduce the reservoir of infection (King et al., 2007). In 2011 the UK Government decided to permit low cost badger culls in some regions of England (DEFRA, 2011). The first such badger culls took place in 2013.

TB is a complex disease with dynamic coupling of infection both within and between at least two species – in the UK badgers and cattle (Krebs et al., 1997). Recent genetic evidence has demonstrated the persistence of TB strains on or near farms for several years despite clear whole herd tests over the same period with ongoing infection in the local badger population with strains closely related to those that had been seen on the farms (Biek et al., 2012). While this is compelling evidence that the same strains exist in sympatric cattle and badgers, it is equally as supportive of badgers transmitting TB to cattle as it is of cattle transmitting TB to badgers. There is evidence that both cattle and badgers transmit TB to individuals of their own species; there is experimental evidence that cattle-to-cattle transmission of TB occurs (Goodchild and Clifton-Hadley, 2001), and that TB is transmitted both among and between social groups of badgers (Goodchild et al., 2012). Cattle transmit TB to each other even at early stages of infection (Kao et al., 2007) as well as to badgers (Woodroffe et al., 2005). There is a long established programme of cattle testing against TB with immediate slaughter of animals giving a positive test, which should remove infective animals from the population. Despite this, a significant number of cases of TB are first detected at slaughter (Gibbens, 2009).

It is likely that TB circulates within both the badger and the cattle populations, as well as between the two species; and that a fraction of cattle infected with TB test falsely negative on the standard tests and therefore persist in the herd for long enough potentially to transmit the disease to other cattle and to badgers. The standard test for TB has a reported median accuracy of 80% and thus will be expected to miss approximately two out of every ten TB infected cattle (DEFRA, 2009). A recent analysis has demonstrated that if the common heminth, *Fasciola hepatica*, is present then the likelihood of detecting TB, if it is present, in a herd drops by about a third (Claridge et al., 2012). This suggests that at least in areas where this parasite is abundant, such as in the west of the UK (Claridge et al., 2012), the accuracy of the standard TB test will be reduced below, possibly substantially below, 80%. These observations are consistent with the fact that approximately 25% of TB cases are first identified at slaughter (Gibbens, 2009), i.e. some TB infected cattle remain in the herd because they are not detected by the tests used to detect the disease. The existence of cattle-to-cattle transmission of TB and a failure to detect an animal as infected with TB when it does harbour the disease are both necessary to explain the observation that the spread of TB in the cattle population is better described by cattle movement data than by other factors (Gilbert et al., 2005), i.e. cattle which were believed to be clear of TB are in fact harbouring the disease and move it to new areas when they are moved between farms.

The task of quantifying and modelling dynamics of TB between and within badgers and cattle is not trivial (Godfray et al., 2013). While statistical and analytical methods provide a rigorous way of analysing existing data, they are often poor at predicting scenarios and a computational approach is likely to be more desirable (Evans et al., 2013a). Computational models provide a valid alternative to expensive experimental approaches (Godfray et al., 2004) as a method of testing the likely effects of various strategies designed to control or eradicate TB in cattle. To be useful such models need to represent the modelled system in sufficient detail to allow realistic predictions to be made about the outcome of any control strategy (Evans et al., 2013b). Individual and Agent based models (IBMs) link individuals with populations through fitness maximisation, energetics, and behavioural decisions of individuals that influence population-level outcomes (Evans et al., 2013a; Moustakas and Evans, 2013; Moustakas and Silvert, 2011; Zhang et al., 2014). Model coupling (Verdin et al., 2014) of multiple dynamically acting animals can provide powerful predictive tools (Evans et al., 2013a).

We have developed a computational model of TB that includes both badgers and cattle with details of both the natural history of badgers and the husbandry of cattle. We have used this model to explore the effects of different potential control strategies (badger culling, changing the TB testing interval and accuracy), and also to investigate spatio-temporal epidemiological effects (Christakos and Hristopulos, 1998; Lin et al., 2014) such as the effects of cattle movements, the rates of transmission both between and within the two species, the effects of farm (and hence herd) sizes as well as winter housing (bringing all the cattle on a farm into one area during the winter months). This model is run on grid of cells such that cell sizes scale to the mean badger territory area, badger culling and culled induced migration areas, farm area, cattle movement distances between farms, cattle winter housing, and the scale of cattle movements between farms.

**Methods**
*Overview*

The model is spatially explicit and is run on a lattice of cells, it is based on agent-based models following in detail the annual cycle of individuals of two agents, badgers and cattle. Model description is comprised by two sections: A description of the model is provided here while technical descriptions, the rationale behind each model section and a detailed referenced parameter space explored is provided in Supp. 1. A schematic representation of the model is shown in Fig. 1. In order to account for dynamics of TB within and between badgers and cattle, we have explored N=465 simulation scenarios in total; $N_1$=157 simulation scenarios with badgers and cattle, $N_2$=154 simulations scenarios with cattle but without badgers, and $N_3$=154 simulation scenarios with badgers but without cattle. An explicit list of all model input variables and the parameter space explored for each variable in simulation scenarios performed is provided in Supp. 2.

*Model description*
*Grid*

It has been reported that on average badgers spend more than 95% of their time in their group territory (Roper and Lüps, 1993). Further around 80% of the population are reported to be within their natal group four years after their birth(Woodroffe et al., 2005). In general, badgers usually do not move from their home range area if population sizes are constant and in the absence of culling (Riordan et al., 2011). We thus decided to set the mean cell size to be equal to the mean badger home range area in the UK, which is reported to be 0.7 km$^2$ (Cheeseman et al., 1981; Krebs et al., 1997) corresponding to 0.84 km on a side of each cell. Having determined the surface area of each cell the initial number of badgers and cattle included in the simulation was determined using published data of the mean number of badgers and mean number of cattle per km$^2$.

Grid: The total simulation area covers 16384 cells x 0.7 km$^2$ cell$^{-1}$ = 11,468.8 km$^2$. The cell size of 0.7 km$^2$ cell$^{-1}$ was chosen in order to correspond to the mean home range area in the UK of a group of badgers (see 'badger movement'). The model is run on a lattice of 128 x 128 cells (totalling 16384 cells). No periodic boundary conditions were employed. Farms are initially distributed on the grid with farm sizes consisting of a number of cells forming blocks. The eight cells adjacent to the current cell are considered as neighbouring cells on each time step.

Badgers: Badger density is determined by the availability of food and settlement sites (Kruuk and Parish, 1982). In the UK badger density is known to vary with latitude and it is reported to be higher in the south than in the north (Table 3.3 in (Krebs et al., 1997)). We chose to initialise the model with 5.7 adult badgers per km$^2$ a value corresponding to Bristol area (Table 3.3 in (Krebs et al., 1997)). This resulted in 5.7 badgers km$^{-2}$ x 0.7 km cell$^{-1}$ x 16384 cells = 65,372 badgers. Each badger was followed individually throughout the simulation period on each simulation scenario.

Cattle: Cattle data followed dairy cattle whenever data were available, while when data were not available for dairy cattle explicitly we used data that did not distinguish between beef and dairy cattle. According to (Eurostat, 2009) the mean dairy farm size in the UK is 75 ha (= 0.75 km$^2$), and the mean number of cattle per farm is 91, corresponding to 1.21 cattle per ha and thus 121 cattle per km$^2$. As cell size on the grid is 0.7 km$^2$ this corresponds to 84.7 cattle per cell and in total 16384 cells x 84.7 cattle cell$^{-1}$= 1,387,725 cattle. Each cattle was followed individually throughout the simulation period on each simulation scenario.

All simulation scenarios were replicated three times, (i) with badgers and cattle with population numbers as described above, (ii) with all other parameters been identical but without badgers, (iii) all else been equal but without cattle, in order to investigate dynamics of TB within and between the cattle and badger population. An exhaustive number of simulation runs with all parameters space explored here would require over 600,000 simulations which are very computationally demanding. We applied the input sensitivity analysis method Latin Hypercube sampling to save processing time while covering as much parameter space as possible (McKay et al., 1979). Latin hypercube sampling generates an unbiased parameter subset of the entire parameter space and it is commonly employed in several disciplines to analyse large computationally intensive datasets (Marrel et al., 2014). We included all the variables of the model and the feasible values for each variable (listed in Supp. 2). For each parameter, 11 values covering the whole range of the parameter space (defined by the min and max value of the variable, values specified in 'parameter space and rationale' in Supp. 1 for each variable) were specified and reordered randomly generating 10 input parameter sets (Meyer et al., 2007). This procedure was replicated three times. The model was run 10 times for each set of parameters (Meyer et al., 2007). The simulation length during this process was set to 222 time steps (months) equal to the time span of the publicly available data (Jan 1996 – June 2014) regarding infected cattle and total cattle tests (DEFRA, 2014a, c).As the data were on a different scale than the model we analysed percentage of infected cattle by dividing infected cattle with total cattle tests. This is a crude estimate as infected herds are tested more often, there are geographic differences in testing frequency and other biases (DEFRA, 2014a). Infected cattle in the model are cattle that have been detected infected (i.e. cattle that the test detected infected) and total cattle tests include all cattle tests in each time step including regular tests and pre-movement tests. The resulting % of infected cattle was averaged over time and over the simulation runs. After a successful check of normality and independence of errors, a linear regression model was applied with percentage of infected cattle as response variable and all the input variables as explanatory variables (Meyer et al., 2007). To obtain a measure for the relative importance of the remaining parameters with respect to % of infected cattle, the standardized regression coefficients were calculated as the absolute ratio of the coefficient and the corresponding standard error (Meyer et al., 2007). This process was repeated twice, one with badgers and cattle in the model and a second with cattle only. As deduced by Latin hypercube sampling for (i) badgers and cattle $N_1 = 157$ simulations were required, (ii) for cattle only $N_2 = 154$. We were unable to conduct this analysis for badgers only as we had no access to a badger time series dataset. We arbitrarily used an equal number of simulations for the badgers only (so $N_3 = N_2 = 154$) simulations as for the cattle only simulations. The standardised regression coefficients derived from cattle only were used for badgers only simulations too. Thus 465 simulations in total were explored with each simulation being replicated 10 times. Observed and simulated average values were compared with t-tests with Welch correction (Welch, 1947) for unequal variances (Meyer et al., 2007).

*Time step*
All parameters related to time are defined on an annual basis. The model time step was set to one month. Time step was set to one month as there was no biological process that we could identify in badgers and cattle acting in a time frame shorter than a month. Each simulation scenario was run for 30 years (360 months).

*Badgers*
*Badger demographics*
Badgers demographics are defined by a birth and death rate as well as a maximum life expectancy that differs between healthy and infected individuals. Birth rates are implemented deterministically while death rates are stochastic, and both vary ±2% between years from the input value. Birth occurs only in spring months while death continues throughout the year (Neal and Harbison, 1958). In our model badgers reproduce clonally but accounting for a 50:50 % ratio of male to female individuals within the population: Each year only (the female) half of the population may give birth to offspring. Individuals that give birth produce from one to five (mean = three) cubs per litter (Byrne et al., 2012). Badgers live according to a maximum life expectancy when healthy and a reduced maximum life expectancy when infected (Little et al., 1982). Deaths of infected badgers are not counted in the annual death rate of healthy badgers (i.e. the two death rates are different). Deaths are implemented as following: Badgers from the oldest age-class die during the year. If there are insufficient healthy or infected badgers in the oldest age-class to fulfil the relevant annual death rate for the population as a whole, or infected badgers, badgers from the next oldest age-class are killed randomly in space until the badger death rate for total, or infected badgers is fulfilled for that year. If during current year after fulfilling the annual death rates for healthy and infected badgers some badgers remain in the oldest age class for either healthy, or infected badgers, these individuals are killed and so the death rate during that year will be higher than the typical annual death rate. Badgers located on culling cells suffer culling mortalities in scenarios during which culling is applied (Smith et al., 2012). Culling mortalities are not part of the death rate of badgers i.e. culling mortality is additive to death rate.

*Badger movement*
For each time step 95% of badgers within each group do not move while the remaining 5% move (Roper and Lüps, 1993). The 5% of moving badgers will move to the 8 neighbouring cells and return back on current cell where their home group is located during the next time step. When in one of the 8 neighbouring cells, the moving badgers can both infect and get infected by badgers and/or cattle of the neighbouring cell. Badgers also move between cells based on minimum and maximum group sizes (Cheeseman et al., 1988; Kruuk and Parish, 1982), and culling (Woodroffe et al., 2006b); If the number of badgers on the current cell is smaller than the maximum number of badgers per cell, newly born badgers stay on current cell. Newly born badgers move to other cells to form groups when the number of badgers on the current cell is larger than the maximum number of badgers per group size (Rogers et al., 1998; Roper et al., 2003) a year after their birth (i.e. not during the year that were born). These solitary badgers keep on moving until they find a group populated with lower number of badgers than the maximum group size or until they encounter another solitary badger and form a new group. A detailed technical description of badger movement implementation is provided in Supp. 1 'badger movement'.

*Culling*
In scenarios in which culling is applied, it is implemented by removing a proportion of the badgers within a number (block) of cells forming a square. The proportion of badgers removed within a culling block(s) is defined by culling intensity (Wilkinson et al., 2009). The model assumes that culling-induced migration happens always as long as there is a single badger killed by culling in the badger group (Bielby et al., 2014). Once at least one badger is removed due to culling in culling cells, culling causes badger migration with badgers located in the current cell (located within culled areas) migrating to neighbouring cells (Karolemeas et al., 2012). A full technical description of the implementation of culling is provided in Supp. 1 'Culling'. This is implemented as follows: Badgers move within the next time step away from current cell into a new cell located within the culling-induced migration distance. The culling-induced migration distance is up to four cells from current cell per time step corresponding to a distance close to 7.5 km from their home settlement (Sleeman, 1992). Culling induced migrating badgers seek to form a new group in the neighbourhood of cells located within the maximum migration distance (Riordan et al., 2011).

*Cattle*
*Winter housing and Cattle movement*

Within the farm, cattle can move only on the blocks of cells that comprise the farm. In scenarios in which winter housing is applied, during the winter months all cattle are summoned to one cell of the farm were winter housing takes place (Brennan and Christley, 2012). Note that the cattle-to-cattle infection rate *per se* does not vary whether winter housing is applied or not. In addition badgers from the current cell (coinciding with the cell of winter housing) can enter the farm buildings where winter housing is applied (Tolhurst et al., 2009). During non-winter months (and during all months in scenarios in which winter housing is not applied) cattle perform random walks on farm cells. Cattle are bought from other farm owners and can move between farms on every time step (Gilbert et al., 2005). Cattle movement between farms is defined by two parameters: the percentage of the total population of cattle moving (sold) every year, and the mean distance between the originating farm and the destination farm that the sold cattle move every year (Gilbert et al., 2005). Cattle that move into a new farm are removed from the population of the originating farm and added to the cattle population of the destination farm. New cattle are initially placed on the cell of the new farm where winter housing takes place (central cell) and can random walk within the new farm if no winter housing applies or stays on that cell if the movement into the new farm was made during winter months. Cattle moving between farms (sold) are tested prior to moving into the new farm (AHVLA, 2013), see section 'cattle testing'.

*Cattle demographics*

Cattle demographics are defined by a birth and death rate as well as a maximum life expectancy that differs between healthy and infected individuals. Birth rates are implemented deterministically while death rates are stochastic and both vary ±2% between years from the input value. Reproduction is clonal  but accounting for the ratio of male to female individuals within the population (Thibier and Wagner, 2002) see Supp. 1 'cattle demographics for more details. The proportion of the total population of cattle that gives birth each year produces on average one calf. Cattle live according to a maximum life expectancy when healthy (NSF, 2010) and a reduced maximum life expectancy when infected (Costello et al., 1998). Deaths of infected cattle are not counted in the annual death rate of healthy cattle (i.e. the two death rates are different). Deaths are implemented as following: All cattle from the oldest age-class die during that year. If there is insufficient numbers of healthy or infected cattle from the oldest age-class to fulfil the annual death rate for either healthy, or infected cattle, a number of individuals from the next oldest age-class are killed randomly in space until the cattle death rate for healthy and infected cattle is fulfilled for that year. If during the current year there are more cattle in the maximum oldest age class of either healthy, or infected cattle than are required to fulfil the annual death rate these individuals are killed and death rate during that year will be higher than the typical annual death rate.

*Cattle testing*

Cattle are tested with a frequency defined by an interval of months based on the initial cattle infection ratio (DEFRA, 2014a); (e.g. every 48 months  - 4 years - when the initial % of infected cattle was < 0.2%). The cattle test is characterised by a testing accuracy defined as a percentage of false negative detections (Szmaragd et al., 2012); (i.e. a testing accuracy of 80% means that for every 100 infected cattle tested, 80 cattle will be detected infected). Detected infected cattle are slaughtered and removed from the population. Cattle that move between farms are tested prior to moving with the same testing accuracy. Cattle that are infected, and detected while moving into a new farm are slaughtered and removed from the population (AHVLA, 2013). Cattle moving onto a new farm that are tested, and found to be uninfected, or are infected but are undetected are moved onto the new farm.

*Infections*
*Badger to cattle*
Badgers infect cattle located on current cell (Benham and Broom, 1991; Smith et al., 2001) with a probability defined by the badger-to-cattle infection rate. The implementation is stochastic & density-dependent (Wilkinson et al., 2009): For every infected badger on the current cell let the infection rate be a number $I_{b2c}$ and $rnd$ a uniform random number drawn in [0, 1]. If $rnd \leq I_{b2c}$ the infected badger will infect a cow on current cell. The process is repeated during every time step for every cell for every infected badger with every non-infected cattle individual on current cell. The time of infection is recorded for every cow. The mean life expectancy of newly infected cattle is adjusted from mean healthy life expectancy to mean infected cattle life expectancy from the time of infection.

*Badger to badger*
Badgers infect other badgers located on current cell (Jenkins et al., 2012) with a probability defined by the badger to badger infection rate. The model has the option for different infection rates between and within badger groups however this option was not explored here see 'badger to cattle' section and full technical implementation and rationale in Supp. 1.

*Cattle to badger*
Cattle infect badgers located on the current cell (Woodroffe et al., 2006b; Woodroffe et al., 2005) with a probability defined by the cattle-to-badger infection rate. The process is implemented as described in 'badger to cattle' section and full technical implementation is provided in Supp. 1 'cattle to badger'.

*Cattle to cattle*
Cattle infect other cattle located on their current cell (Conlan et al., 2012; Goodchild and Clifton-Hadley, 2001) with a probability defined by the cattle-to-cattle infection rate. The process is implemented as described in 'badger to cattle' section and full technical implementation is provided in Supp. 1 'cattle to cattle'.

*Initialisation*
The number of farms is calculated by dividing the total number of cells on the grid by the mean farm size (Goodchild and Clifton-Hadley, 2001). Farm structure is implemented on the grid in blocks of cells equal to the mean farm size. Cattle are initially uniformly distributed over the farms by dividing the initial number of cattle by the number of farms. Initially, infected cattle are randomly distributed over the farms. In scenarios in which farm size is larger than one cell (farm sizes are identical within each scenario but they vary between scenarios with sizes spanning from one to four cells corresponding to 70 – 280 ha) the cell closest to the upper left corner of the farm is selected as a winter housing cell location (Brennan and Christley, 2012). Badger groups (setts) are initially randomly distributed on the grid by dividing the total initial number of badgers by the max badger group size defining the initial badger groups on the grid. Note that farm sizes vary from one to four cells while badger home range is always one cell and depending on the initial number of badgers there are several cells without a group of badgers. In the scenarios explored here the initial number of badgers is 65,372 divided by 21 = 3,113 initial badger groups. Thus only 3,113 cells out of 16,384 cells contain badgers initially. Infected badgers are initially distributed on the grid in an aggregated way defined by the percentage (%) of cells containing at least one infected badger and the percentage (%) of badgers from the total initial badger population been initially infected (Delahay et al., 2000). In scenarios in which culling applies, culling takes place in a number of culling areas (Bourne et al., 2007) defined by the upper left and lower right coordinate of the cells defining a square block of cells.

Linear mixed effects models were used in order to conduct sensitivity analysis of model outputs between the number of infected cattle (dependent variable), and all the input parameters explored as independent variables (Pinheiro and Bates, 2000). The total number of cattle was used as a random effect in the model in order to account for potential larger numbers of infected cattle in large sized herds that could bias the analysis towards larger herd sizes. The most parsimonious model structure was selected using the Akaike Information Criterion (AIC). Sequentially, we implemented hierarchical variance partitioning (Matsinos et al., 2011) of the covariates of the most parsimonious mixed effects model to account for the contribution of each explanatory variable to the total variance of infected cattle. For details in statistical analysis see Supp. 1.

*Confronting model outputs with data*

We have identified three potential patterns to compare model outputs with data: (i) Effects of culling over time in TB prevalence in cattle within culling areas. (ii) Population increase of badgers over time (iii) Compare model outputs with published studies regarding the main results derived. Details on these three modelling scenarios are provided in Supp. 1 'Confronting model outputs with data'.

**Results**

Sensitivity analysis of model input parameters (Latin hypercube sampling) indicated that observed and simulated values in terms of percentage of infected cattle were not significantly different from each other (t-test statistic = -1.26, P = 0.45).ANOVA results of the most parsimonious mixed model with the number of infected cattle as a dependent variable, show that there were significant effects of the percentage of cattle that are moved in a year ($F_{4, 156}$ = 54.62, P < 0.0001), the distance which cattle were moved ($F_{4, 156}$ = 7.74, P = 0.006), both the cattle-to-badger and badger-to-cattle infection rates ($F_{3, 156}$ = 4.46, P = 0.036; $F_{3, 156}$ = 8.59, P < 0.004), the inter-test interval ($F_{4, 156}$ = 59.80, P < 0.0001) and the accuracy of the test ($F_{3, 156}$ = 3.81, P = 0.053), badger culling ($F_{2, 156}$ = 8.91, P = 0.003) and the initial number of infected badgers ($F_{2, 156}$ = 16.08, P = 0.0001). These results are shown in detail in Table 1 of Supp. 1 and the text therein. Of these effects the ones that explain the most variance (>10% of the total variance) in the number of infected cattle are: percentage of cattle moved – as a greater proportion of the cattle are moved the number of infected cattle increases; the frequency of TB testing, which results in more infected animals being detected when the inter-test interval is short; and the badger-to-cattle infection rate. Of the remaining factors, badger culling explained about 5%, while test accuracy explained less than 3% of the variance in the number of infected cattle (Fig. 2). Of the significant factors only badger culling, cattle testing and controls on cattle movement are potential TB control strategies.

We investigated some extreme scenarios in which all the badgers in the region had been eradicated completely – this might be seen as the logical endpoint of a culling strategy, or the situation that might be found in the centre of a large badger culling zone. In this case we found that the significant independent variables were mean farm size ($F_{2, 153}$ = 5.47, P = 0.021), whether winter housing was practised ($F_{1, 153}$ = 36.26, P < 0.0001) (Fig. 3), the inter-test interval ($F_{4, 153}$ = 36.73, P < 0.0001), the percentage of cattle that are moved ($F_{4, 153}$ = 82.73, P < 0.0001) and the distance that cattle move ($F_{4, 153}$ = 26.15, P < 0.0001). These results are shown in detail in Table 2 of Supp. 1 and text therein. For results regarding simulations following badgers only (no cattle) see Table 3 of Supp. 1 and text therein.

Confronting model outputs with data showed that the model outputs for infections in cattle within culling areas after 5 y were slightly lower than the data (Fig. 4i) potentially due to the fact that culling area was modelled as a square rather than the circle that is implemented in reality (and so the model had higher edge effects). Model outputs in terms of changes in the size of the badger population over time showed values very close to the actual population increase observed in the Bristol area and values between the actual population increases seen in Wytham (Oxfordshire) and in Gloucestershire areas (Fig 4ii). However model outputs were at least three-fold lower than the average badger population increase in several areas in the UK recorded by (Wilson et al., 1997) that suggested a 77% increase in the population of badgers in the UK. A scenario in which the impact of the FMD epidemic on TB was mimicked including the interruption in cattle testing interruption and movement restrictions showed similar patterns to the available data (Fig 4iii).

All scenarios leading to TB eradication in cattle included cattle testing frequency of at least once per year (we explored testing frequency of up to 6 months). Time series model outputs of the monthly number of infected cattle in the population are provided in Fig. 5. These outputs were generated with parameter space as in lines 13-21 in Supp. 2. These five scenarios were run for a fifty-year period from ten years into a run (to exclude initial transitional effects). The outputs included the following scenarios: (1) badgers not culled, cattle tested every year; (2) badgers culled + cattle tested every year; (3) badgers eradicated completely + cattle tested every year; (4) badgers culled + cattle tested every six months; (5) badgers not culled + cattle tested every six months (Fig. 5). Testing accuracy was 80% (Supp. 2).

**Discussion**

*Model complexity & uncertainty*

We have developed a reasonably detailed computational model to follow TB transmission between cattle and badgers by coupling IBMs of the two species. TB is a complex disease (Godfray et al., 2013) and complex questions merit models of levels of complexity similar to the problem addressed (Evans et al., 2014; Evans et al., 2013b). In doing so we are facing a number of challenges: As the number of model components and ultimately model input parameters increases, it is difficult to perform a sensitivity analysis not due simply to the computational time, but mainly to the non-linearities arising from coupling of two animals (Bithell and Brasington, 2009). This is further stressed by the fact that several parameters have rarely been measured. For example the cattle-to-badger infection rate is, as far as we can determine, unknown because most studies quantifying infection rates have focused on cattle infecting cattle (Conlan et al., 2012) or badgers infecting cattle (Benham and Broom, 1991; Smith et al., 2001) but not cattle infecting badgers. Studies providing strong inference (Woodroffe et al., 2006b; Woodroffe et al., 2005) that cattle are infecting badgers do not provide infection rates. We thus do not pretend that all results derived in simulation scenarios here are necessarily real case scenarios. However we have explored computationally a  large number of scenarios (465) derived from published parameter space and thus some combinations of this parameter space are likely to have some merit in reality.

*Confronting model outputs with data & published studies*

The current concern about TB stems from the increase in its incidence and geographical spread since the late 1970s. It is true that the badger population appears to have increased during this period (Harris and Yalden, 2008) but farm sizes have also increased and the practice of winter housing is more common now than it was in the 1970s (Goodchild and Clifton-Hadley, 2001). Model outputs are in agreement with these observations and suggest that all these factors could contribute to increasing the incidence of TB in cattle.  Model outputs are also in agreement with reports that the false negative rate of the standard TB test (Claridge et al., 2012), combined with relatively infrequent testing (Krebs et al., 1997) allows cattle with TB to persist in the herd and pass their infection to other cattle. Testing animals that move between farms is a good step and probably reduces the spread of the disease but it must be recognised that the false negative rate of the test will mean that some of these animals continue to pose an infection risk to the herds into which they move. The variance partitioning results derived here are consistent with (Gilbert et al., 2005) who found that cattle movements explained the spread of TB more completely than any other environmental or anthropogenic factor. In addition we found that winter housing accounts for a large percentage of cattle-to-cattle TB transmission as discussed (but not quantified) in recent reports (Bourne et al., 2007; Krebs et al., 1997). To provide a test of the ability of our model to mimic the result of an event that is known to have changed the dynamics of TB in cattle, we explored a simulation scenario where initial conditions were as in line 9 of Supp. 2 and after 20 simulated years the model's output in terms of total and infected cattle and badgers was used as input to a simulation scenario in which all else was the same as the previous scenario but cattle testing was interrupted for one year as occurred during the 2002 FMD epidemic in the UK (Fig. 4).  This simulation shows that the brief stop in cattle testing which occurred during the FMD epidemic would have increased the number of infected cattle by about 33%, which is comparable with published data (DEFRA, 2014b). Note that we are comparing here the pattern of infections and not the actual numbers, because the data cannot account for cattle that are infected but not detected and model and data differ in scale; We also note that the data are at a national level while the model is simulating the equivalent of a county, and spatio-temporal regional differences could be important but are not considered here (Christakos and Hristopulos, 1998). This figure was produced with a single simulation run and not averaged across a large parameter space as was done in results presented in Figures 2 & 3 and thus there is more uncertainty in parameter space than for the other results presented. The patterns we observed was not sensitive to test accuracy: In a simulation using identical parameters space with the one used to produce Fig. 4iiia but with test accuracy set to 50% a similar trend was produced (Fig. 6). Note that while the pattern of time series is similar the number of (detected) infected cattle was lower when testing accuracy was set to 50%.

*Models on TB & model coupling*

Several models have been built and employed for TB but despite those efforts, models have not yet facilitated informed decisions towards TB eradication – for a review and criticism on models until 1997 see also section 4.5 in (Krebs et al., 1997). Several models built after 1997 have either included only badgers (Shirley et al., 2003; Smith and Cheeseman, 2002) or only cattle (Green et al., 2008). Such models could account for TB transmission within each species but not between species (see also Annex 3 in (King et al., 2007)). Other models included badgers and cattle but cattle did not move between farms, did not infect other cattle, and did not infect badgers (Smith et al., 2001; Wilkinson et al., 2004) - these models did include both badgers and cattle but the coupling was not dynamic: badgers could infect cattle and other badgers, but could not be infected by cattle, and cattle could not infect each other, in addition cattle movement was not included. More recent papers have included cattle movement, cattle-to-cattle, cattle-to-badger infections, and cattle testing (Smith et al., 2012; Wilkinson et al., 2009). However this model was run on a 100 x 100 grid each cell representing just 200 x 200 m, giving a small total grid area of 400 km$^2$, using also periodic boundary conditions (the grid was wrapped to form a torus to eliminate edge effects) (Wilkinson et al., 2009). Furthermore culling was applied by assuming if at least 10% of a social group territory was accessible for culling, then badgers could be removed at the same rate as if 100% was accessible (Wilkinson et al., 2009). We argue that spatial scale is highly critical when coupling models. First the size of the simulation box must be large enough to prevent periodic artefacts from occurring due to the topology of the simulation. In a cell that is too small, an individual may interact with its own image in a neighbouring cell: the model simulates cattle that are almost exclusively infecting cattle of the same herd and farm. A cell size of 200 x 200 m is far too small to provide a model replicate for cattle in a larger scale such as a county or a country. The mean farm size in the UK varies from < 10 ha to > 200 ha per farm (see 'Initialisation' in Supp. 1 and references therein). In a grid that is too small cattle movement between farms, (mean distance travelled per moving individual in the UK was 75.6 km year$^{-1}$ with a standard deviation of 84.13 km year$^{-1}$ (Gilbert et al., 2005)), and culling that needs to span over at least an area of 141km$^2$ (Godfray et al., 2004; Jenkins et al., 2010) cannot be scaled and thus accounted for. We propose that model coupling need not have the same cell size for each life form but it needs to scale at the minimum area requirements for cattle and farms on the one side and badgers on the other. On the other hand grid size needs to be large enough to allow for large-scale phenomena such as cattle movement and culling. Using periodic boundary conditions in a small grid is based in the silent presupposition that a land patch equal to the grid size is representative of the actual situation. We argue that processes such as cattle movement or culling are acting on large scales and thus require a larger total surface area modelled. For a view on scaling issues in gridded models and model structure with scenario boundary conditions see also (Bithell and Macmillan, 2007; Millington et al., 2011). As far as we are aware this is the first model to include the effects of winter housing, and the first 'predictive' TB model (i.e. a model calibrated with data that can be used to investigate hypothetical scenarios (Evans et al., 2013a)) in which model outputs are compared with data.

*Main model outputs- Conclusions*

Our results imply that of all the factors explored here the ones that explain the most variance (>10% of the total variance) in the number of infected cattle are: cattle movement; the frequency of TB testing; and the badger-to-cattle infection rate. Badger culling explained about 5% of the variance in the number of infected cattle while test accuracy explained less than 3% (Fig. 2). Of the significant factors only badger culling, cattle testing and cattle movement are potential TB control strategies.

We investigated an extreme scenario in which all the badgers in the region had been eradicated completely and there were no badgers present. In this case we found that the significant independent variables were mean farm size, whether winter housing was practised, the inter-test interval, the percentage of cattle that are moved and the distance that cattle move. In particular the interplay between winter housing and mean farm size (Fig. 3) has a predominant role on the spread of the cattle-to-cattle disease dynamics. When winter housing is practised there is considerably higher disease spread (as there is a higher density of cattle and contact rates become higher) and this becomes more pronounced with increasing farm size as more cattle are brought within the same winter housing facility and the number of animals that any one individual is in contact with rises. A potential way to surmount this problem would be to have several winter housing units with increasing farm sizes in order to spread the cattle population during winter months. It would be easier to eradicate TB from ten herds each of 30 cattle than from one herd of 300.

Culling of badgers does seem to be a strategy that will eventually lead to a lower incidence of TB in cattle but does not seem capable of eradicating the disease – in our scenarios in which badgers had been removed completely TB remained endemic in the cattle herd even with regular testing and testing on movement (Fig. 2 & Fig. 5). A strategy that removes TB more quickly is frequent testing of cattle, which was equally true whether badgers were culled or unculled or even absent, TB declined in cattle more rapidly with regular six monthly tests than with any other strategy for TB control explored here. In fact there was almost no difference between the number of infected cattle in a scenario in which badgers were culled or not culled as long as testing was sufficiently frequent (Fig. 5). This is result is valid even if the test has imperfect detection accuracy. Thus, our results imply that more frequent testing would facilitate the eradication of the disease despite the imperfect test detection. One possible problem with frequent testing would be the creation of immunity to the testing agent, we have assumed that an infected individual is as likely to be detected on test n+1 as on test n.

Our results suggest also that two little studied variables play an important role in the disease: The cattle-to-badger and the badger-to-cattle infection rates. In addition experimental studies should examine if indeed there is a difference in the actual values of badger-to-badger infection rates between and within badger groups (see also Supp. 1 badger-to-badger infection rates). We suggest that experimental field studies either *in situ* or in laboratories should be set up to quantify the infection rates between cattle and badgers, and local trapping experiments to provide geographic variation of the percentage of infected badgers.

**Acknowledgements**

We thank Ian Boyd for comments on an earlier draft, and William Wint for providing us data on cattle movement. We also thank Harry Constantinides for invaluable help during model coding in C#. Comments from two anonymous reviewers considerably improved an earlier manuscript draft.

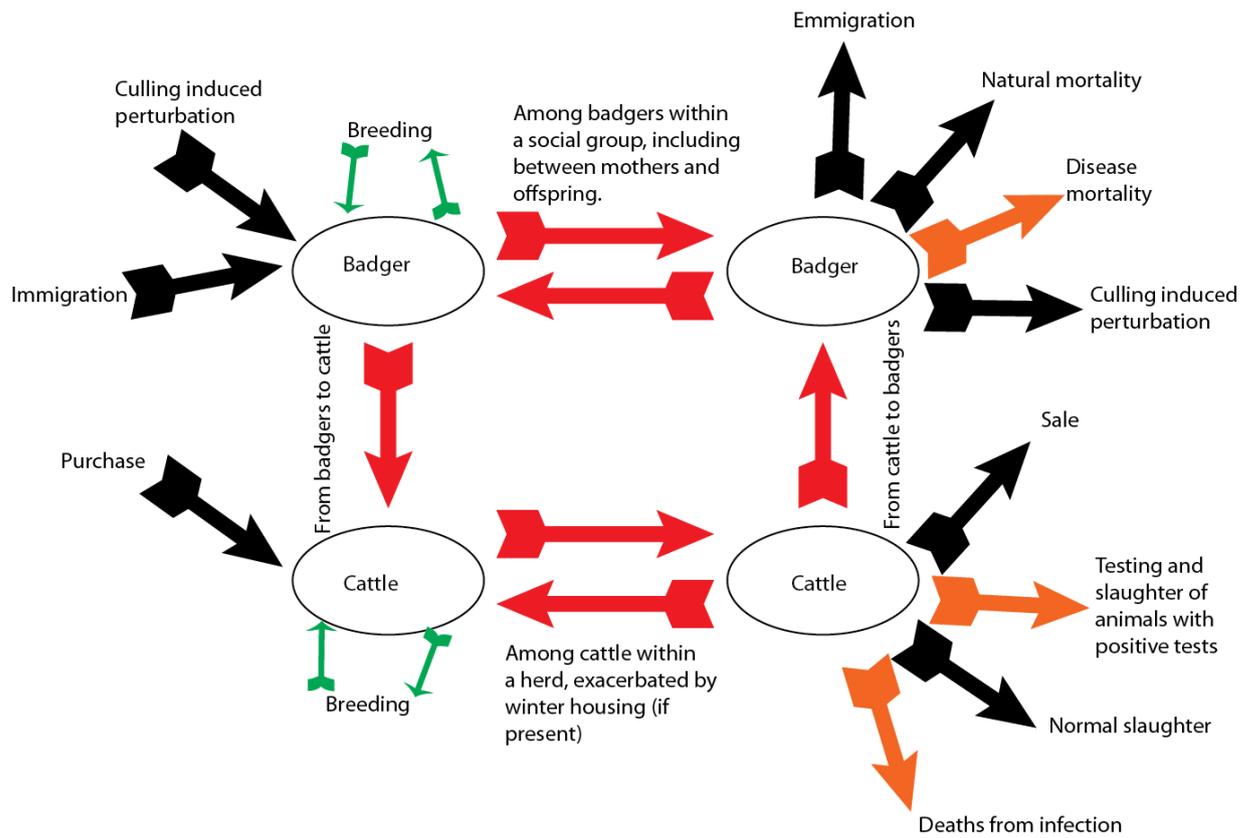

**Figure 1.** A schematic representation of the model following population dynamics of cattle and badgers and the spread of bovine tuberculosis (TB) between and within the two species. Different acting processes are marked with different colours: red is disease transmission, orange are factors that only apply if there is disease, green are demographic processes, and black are demographic processes that will influence the disease.

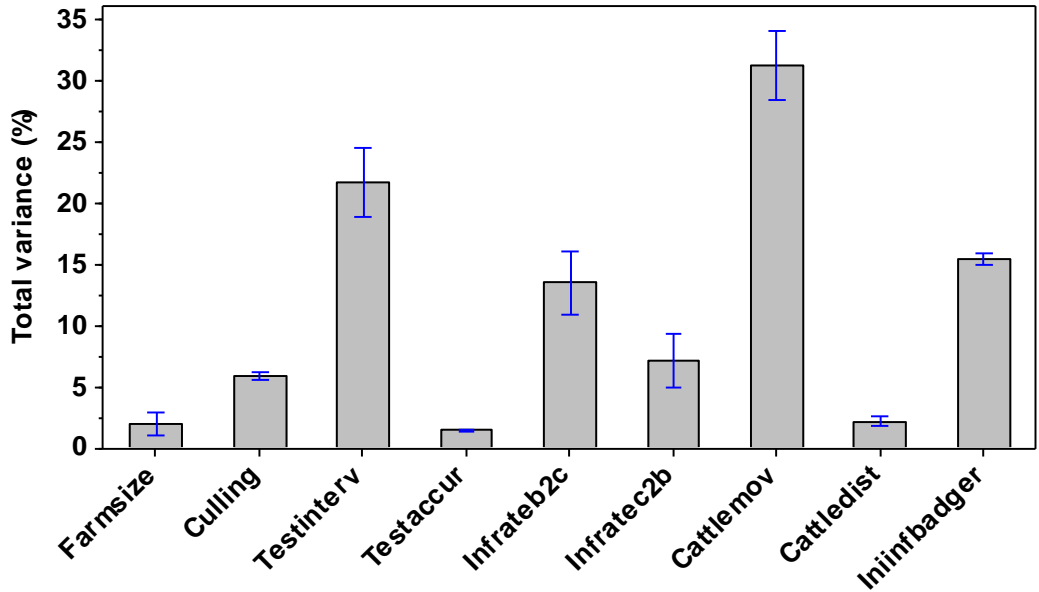

**Figure 2.** The percentage of total variance in the number of infected cattle explained by the significant independent variables in the mixed model ANOVA. Data show the mean percentage (± 95% confidence interval) of the total variance explained by each factor. Mean calculated over 156 simulations (badgers and cattle). Note the total number of cattle was used as a random effect and so the number of infected cattle is unaffected by the total number of cattle in the population.

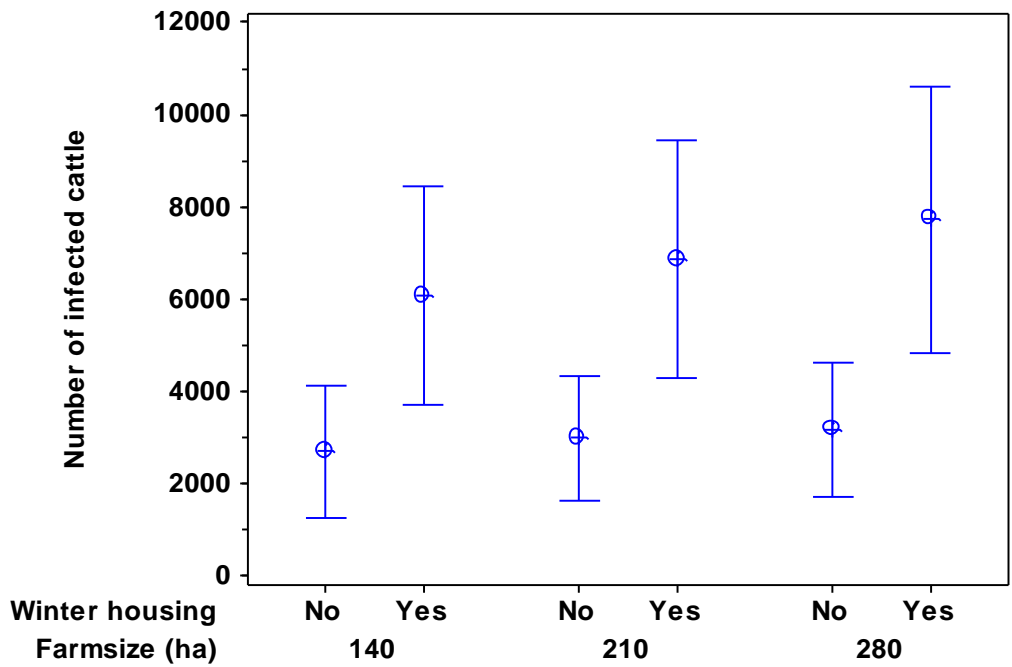

**Figure 3.** Number of infected cattle as a function of winter housing across different farm sizes in the absence of badgers (simulations run with cattle only). The practice of housing all cattle on a farm together over winter increases the number of infected cattle in the population especially when the farms are large. Figure shows the mean and 95% confidence intervals for three farm sizes (140, 210 and 280 Ha) each with and without winter housing.

**i. TB prevalence in cattle after 5 y in proactive culling area**
95% CI for the Mean

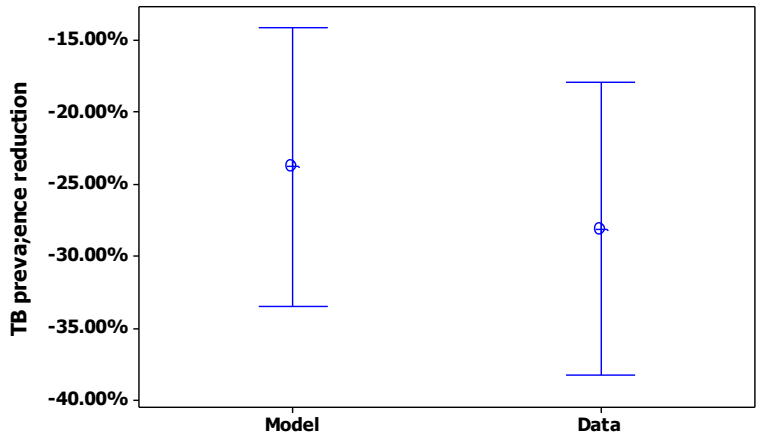

**ii. Percentage of change in the number of badgers**

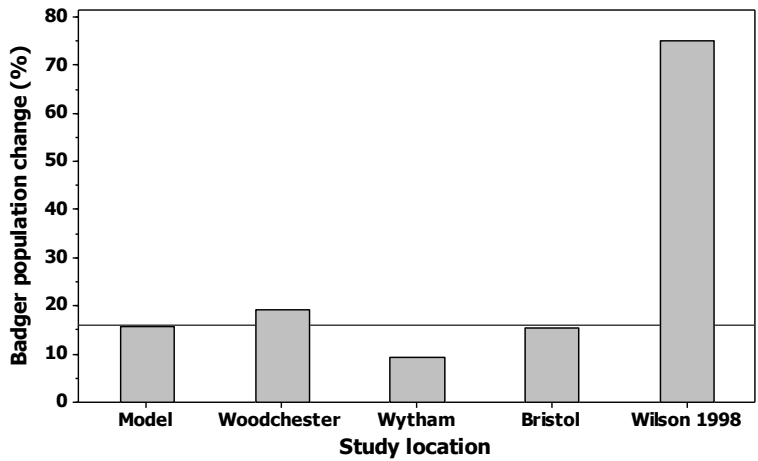

**(iii) a. Testing interrupted during the foot and mouth epidemic - model**

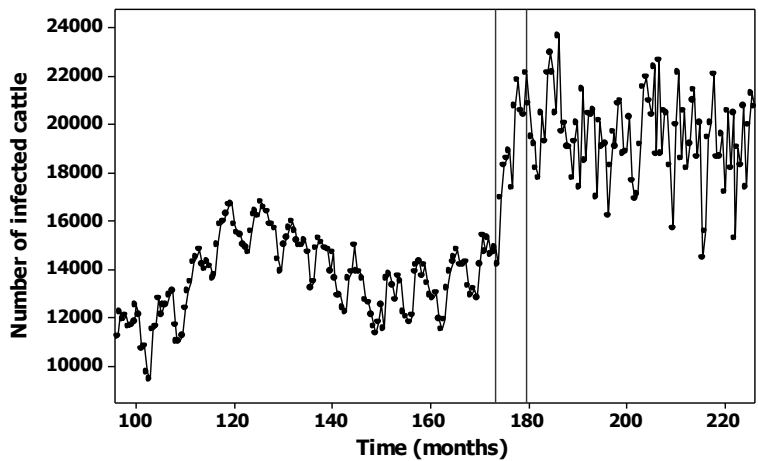

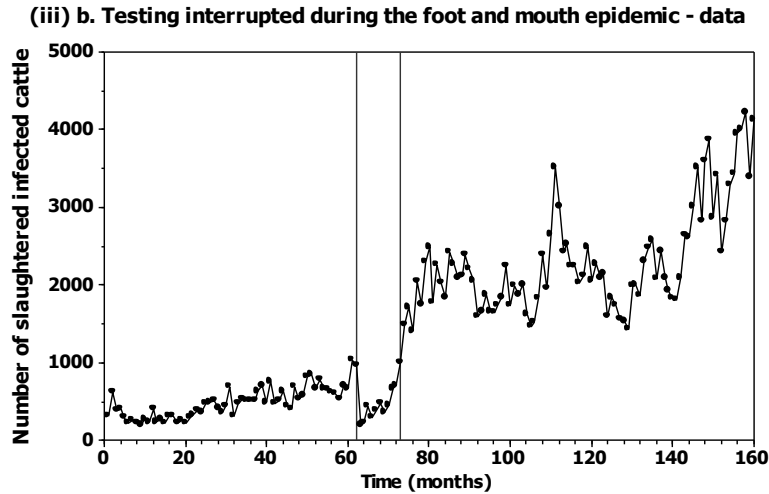

**Figure 4.** Confronting model outputs with three published patterns: **(i)** Impact of five years of culling on badgers within culling areas of 100 km². **(ii)** Percentage of change in the population of badgers over time regarding model's outputs vs. three locations in the UK and the mean of several locations in the UK (Wilson 1997). Reference line shows model output. **(iii). (a)** A scenario simulating TB cattle testing interruption as it occurred in the UK during the foot-and-mouth epidemic**.** Vertical lines indicate the period of cattle testing interruption. **(b)** Time series of TB detected infected cattle slaughtered **-** data from DEFRA on infected cattle during the foot-and-mouth epidemic. Vertical lines indicate the period of cattle testing interruption. For details and references regarding all comparisons between model outputs vs. published data see Supp. 1 'Confronting model outputs with data'.

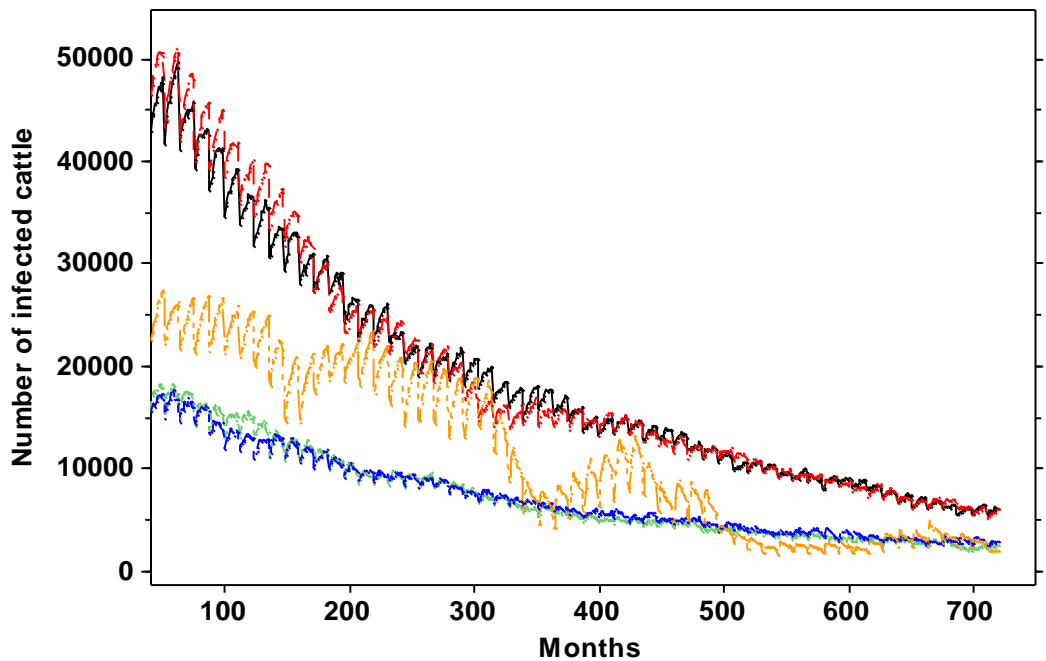

**Figure 5.** Output from five scenarios for a fifty year period running from ten years into a run (to exclude initial transitional effects). Lines show the monthly number of infected cattle in the population. All scenarios are similar and have the same basic parameters for the biology of badgers, the transmission of the disease and the husbandry of cattle with the following differences: Scenario 1 (Black): badgers not culled, cattle tested every year; Scenario 2 (red): badgers culled + cattle tested every year; Scenario 3 (orange): badgers eradicated completely + cattle tested every year; Scenario 4 (green): badgers culled + cattle tested every six months; Scenario 5 (blue): badgers not culled + cattle tested every six months. The parameter space used for generating those time series is listed in Supp. 2, lines 13-21.

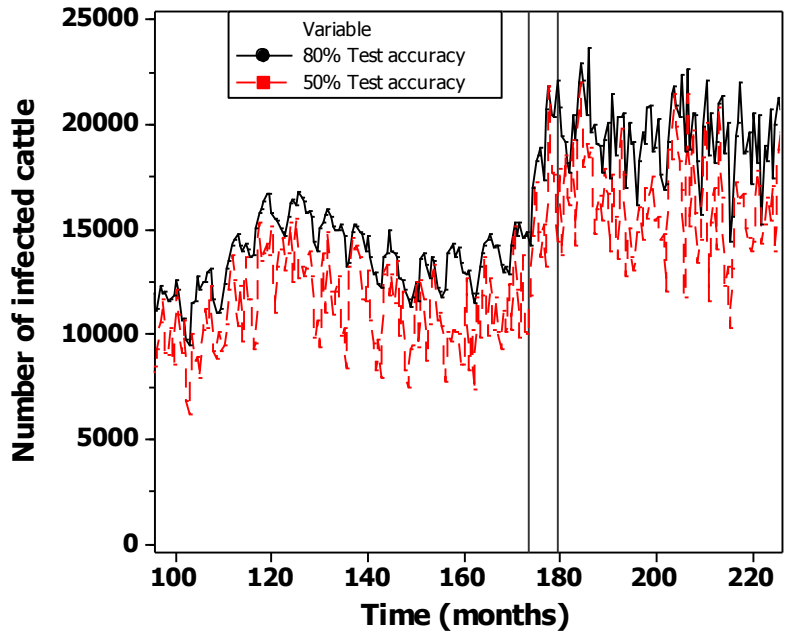

**Figure 6.** Simulating TB cattle testing interruption as it occurred in the UK during the foot-and-mouth epidemic (FMD) with varying testing accuracy**.** Identical parameters space with the one used to generate Fig. 4iiia was employed but testing accuracy was set to 50% instead of 80%. Vertical lines indicate the period of cattle testing interruption.

**Supplement 1**

**Coupling models of cattle and farms with models of badgers for predicting the dynamics of *bovine tuberculosis* (TB)**

**Authors: Aristides Moustakas and Matthew R. Evans**

**Technical model description, rationale behind each model section & parameter space explored**

In this document the rationale behind model building on each model section is explained and the parameter space that was explored is described. Some parts of the model description that were either too technical or too long for the main text are listed in this document under 'Model description' prior to 'Rationale & Parameter space' text of each model section. The model was coded in C#. Simulations were performed on a cluster of computers comprised of 106 machines that each have 12 cores and 24 gigabytes of RAM, and two machines that each have 48 cores and 512 gigabytes of RAM.

***Badgers***

*Badger demographics*

*Rationale & Parameter space*

Badger life expectancy when healthy was set to 5 years as given by Harris and Cresswell (1987) and Cheeseman et al. (1988b) for the Bristol area. Badger life span when infected with TB is known to be less than when not infected, however, although it is known that TB is not necessarily immediately fatal, the time from infection to death is reported to vary between studies spanning from 'a rapid course' up to 709 days (Clifton-Hadley 1993) and up to 3.5 years when in captivity (Little et al. 1982). We thus used the median value (two years) as a predominant value and used also one year and three years in simulation parameter space. Badgers give birth in late winter with mid-February been most common (Cresswel et al. 1992). We have thus used February as badger birth month. Badger population is reported to overall increase in the UK (Wilson et al. 1997). We have used mean annual death rate of 34.5 % (Harris and Cresswell 1987; Cheeseman et al. 1988; Table 3.3 in Krebs et al. 1997) corresponding to Bristol area. Mean annual birth rate for the same area is reported to be 41.4% (Harris and Cresswell 1987; Cheeseman et al. 1988; Table 3.3 in Krebs et al. 1997). However this value is unlikely to hold true for the rest of the UK as this would imply a mean net annual growth rate of 6.9% and thus the badger population in the UK would have doubled in 10 years. (Doubling time is determined by dividing the growth rate into 70. The number 70 comes from the natural log of 2, which is 0.70 and thus 70/6.9 = 10.14 years). While there is evidence that the population of badgers in the UK is increasing (Wilson et al. 1997; Bourne et al. 2007), there are no data verifying specific growth rates at a country scale. We have thus used growth rates in parameter space spanning from 33 to 38 % year$^{-1}$ with increments of 1%, resulting in a net annual growth rate -1% to 3.5%. Death rates were always 34.5%, while birth rates varied accordingly as described.

*Badger movement*

*Model description*

Badger movement is implemented as following: Badgers seek to form groups within the nearest 8 neighbouring cells. If these cells do not fulfil the criteria for forming or joining a group, badgers move to the next-nearest cells (the nearest cells of the 8 neighbouring cells consist of a neighbourhood of 16 cells) and badgers seek to join or form a group. If the criteria for joining or forming a group are still not fulfilled (too few or too many badgers on cell) then badgers seek to form or join a group in the nearest 32 cells. If badgers cannot form or join a group in the nearest 32 cells keep on seeking in the nearest 64 cells until they can for or join a group within the neighbourhood of cells defined by the maximum culling induced migration distance (Woodroffe et al., 1995). Badgers stop seeking to form or join groups of other badgers when the criteria of forming or joining groups cannot be fulfilled, and settle on a cell from the 64-cell neighbourhood other than the current cell with equal probability. This process (seeking for a cell to form or join a group) occurs on the same time step that the badger was born. When culling activities take place on current cell, badgers move to neighbouring cells to form new groups, see section 'culling'.

*Rationale & Parameter space*

Badgers live in social groups (Kruuk and Parish 1982; Cheeseman et al. 1987). Group size is dependent upon prey biomass per unit area (Kruuk and Parish 1982) but there is no correlation between group size and territory size (Kruuk and Parish 1982). In a six-year-study it was reported that a badger home range area may support between two and 21 badgers (Cheeseman et al. 1987; Krebs et al. 1997). We have thus parameterised the model to include as minimum badger group size = 2 badgers and maximum badger group size = 21 badgers. In terms of movement it is hard to distinguish between death and emigration in badgers since carcases are rarely found (Krebs et al. 1997). However, it is known that a proportion of cubs are migrating to form or join different groups (da Silva et al. 1994). Badgers that leave their natal territories typically join neighbouring groups, and seldom move more than 2 km (Cheeseman et al. 1988a; Woodroffe et al. 1995). We have parameterised the model so that badgers can move up to three cells from current cell in any possible direction resulting in a maximum dispersal distance of 3 cells x 0.84 km = 2.54 km.

*Culling*

*Model description*

This is implemented as following: Badgers move within the next time step away from current cell into a new cell located within the culling-induced migration distance. The culling-induced migration distance defines a neighbourhood of near cells (not necessarily only the nearest adjacent cells) where badgers migrate within the next time step. Culling induced migrating badgers seek to form a new group in the neighbourhood of cells located within the max migrating distance (Riordan et al., 2011). This is implemented as follows: Badgers seek to form groups within the nearest 8 neighbouring cells. If these cells do not fulfil the criteria for forming or joining a group, badgers move to the nearest cells of the previous 8 cells (the nearest cells of the 8 neighbouring cells consist of a neighbourhood of 16 cells) and badgers seek to join or form group. If the criteria for joining or forming a group are not fulfilled (too few or too many badgers on cell) then badgers seek to form or join a group in the nearest 32 cells. If badgers cannot form or join a group in the nearest 32 cells keep on seeking in the nearest 64 cells until they can for or join a group within the neighbourhood of cells defined by the maximum culling induced migration distance (Carter et al., 2007). The above described process (seeking a cell to form or join a group) is performed within the next time step after the first culling activity on current cell occurred. Badgers stop seeking to form or join groups of other badgers when the criteria of forming or joining groups cannot be fulfilled, and settle on a cell from the 64-cell neighbourhood other than the current cell with equal probability. If the settled cell (regardless upon whether the badger formed or joined a group or not) is a culling cell, the process is repeated during every time step for a ratio of badgers within the culling block of cells. All culling activities are pro-active i.e. in scenarios that culling is applied, culling takes place in culling areas regardless upon whether there are TB incidents or not.

*Rationale & Parameter space*

Badger culling causes migration and increases badger motility (Carter et al. 2007; Riordan et al. 2011). In a study using telemetry it was reported that when persecuted badgers may move up to 7.5 km from their home settlement (Sleeman 1992) and that this movement has been attributed to badgers being social animals – when group sizes become too small badgers move to other areas seeking to form a group (Woodroffe et al. 1995). We have parameterised the model so that badgers may migrate up to 4 cells from the current cell in order to form or join groups within one time step (month) corresponding to 4 x 0.84 km = 3.36 km a value close to the mean reported by Sleeman (1992). It has been reported that to be 97.5% confident that culling will be beneficial it must be carried out over an area of at least 141 km$^2$ (Jenkins et al. 2010). In the model, in scenarios in which culling was applied culling takes place in blocks of cells comprising of 14 x 14 cells resulting in 196 cells x 0.7 km$^2$ cell$^{-1}$ = 137.2 km$^2$ a value close to the 141 km$^2$ reported by Jenkins et al. (2010). Simulation scenarios included no culling, culling in one block of 14 x 14 cells or culling in three non-adjacent to each other blocks of 14 x 14 cells each. Culling intensity is reported to be an important factor on the total efficacy of culling as a control measure and different culling techniques such as shooting, caging, trapping and gassing have been proposed and applied (Krebs et al. 1997; Wilkinson et al. 2009). Some techniques such as gassing have been reported to be very efficient in removing badgers (p79 in Krebs et al 1997) and have the lowest reoccurrence of TB breakdowns (Table 5.3 in Krebs et al. 1997). We have explored culling intensity of 50, 60, 70, 80, and 90% year$^{-1}$ in culling blocks.

*Cattle*

*Winter housing and Cattle movement*

*Rationale & Parameter space*

Cattle are sold between farms every year and this has been associated with the spread of TB between farms (Gilbert et al. 2005). W. Wint kindly provided the data from 2000 to 2006 for cattle movement between farms in the UK (W. Wint unpublished data). Part of this dataset was used in the Gilbert et al (2005) study. The dataset lists mean and standard deviation values of 'number of events' per year where an event is a case where at least one cattle selling transaction occurred implying that at least one cow was sold and moved between farms as well as the mean and standard deviation values of the distance between the starting and ending farm. The dataset comprises of total cattle (dairy and beef) transactions per year in the UK. Mean distance that cattle moved throughout the UK was 75.6 km year$^{-1}$ corresponding to 75.62 km year$^{-1}$ / 0.84 km ≈ 90 cells year$^{-1}$. The standard deviation of the mean distance that cattle moved in the UK during those years was 84.13 km year$^{-1}$ corresponding to 84.13 km year$^{-1}$ / 0.84 km ≈ 100 cells year$^{-1}$. We thus explored mean annual cattle distance movement parameter space of 0, 30, 60, 90, 120, 150 cells, with 0 distance been a hypothetical scenario with no cattle movement. In the same dataset the mean number of selling transactions per year was 7.5 % implying that at least 7.5% of cattle change farms every year and the standard deviation of that value was 1%. As these numbers describe the mean number of transaction events, the actual number of cattle moving between farms is almost certainly larger we explored a parameter space of 0, 7.5, 8.5, 10, 12 mean percentage (%) number of total cattle moving every year. Value 0 represents a hypothetical scenario with no cattle movement. Winter housing is a common practise in farms (Brennan and Christley, 2012) and takes place during the months that are too cold for grass to grow. We have parameterised the model to apply winter housing from November to April i.e. months 11 to 4 (Food Standards Agency 2007). All simulation scenarios explored using the model have been replicated twice, with winter housing and without winter housing.

*Cattle demographics*

*Rationale & Parameter space*

The majority of dairy cattle in Europe (60% or more) are inseminated using artificial insemination even if a bull is kept on a farm (Thibier and Wagner 2002). It is thus unlikely that a herd needs more than one bull. As mean herd size is 91 cattle per farm in the UK, the sex ratio within each herd is 1/91 = 0.011, or around 1.1% males and 98.9% females. We thus used a sex ratio of 99% females to 1% males within the population. If all females produce one calf per year then the birth rate (assuming that females are 99% of the population) is 99% per year. However half of the calves will be males, and thus killed. Thus, the active population in giving birth to calves that enter the population each year is 99/2=49.5%. However there is a pre-weaning mortality of up to 9.56% (FAO, available at http://www.fao.org/wairdocs/ilri/x5522e/x5522e0a.htm). We have used the median of that value 4.78%≈5%, and thus birth rates are 44.5% per year. Cattle give birth during the spring and we parameterise this to occur in March (month number 3). According to National Farmers Union (2010), 24% of cattle in a herd are culled each year. We thus set as model parameter space of maximum healthy cattle life expectancy of 4 years. Infected cattle seldom live longer than one year (Krebs et al. 1997) and we thus used as maximum infected cattle life expectancy of 1 year. As the population of cattle is essentially controlled by humans (farmers) we have set mean annual birth rates equal to mean annual death rates assuming a constant population all else been equal; However, if for any reason the number of cattle in the herd was reduced, farmers would account for this and adjust cattle population in farm to previous levels by importing cattle or not slaughtering older cattle during that year.

*Cattle testing*

*Rationale & Parameter space*

According to EU Directives 64/432/EEC and 97/12/EC the minimum testing frequency for cattle depends on the percentage of infected cattle herds.

According to the directive annual testing is required unless the percentage of infected herds in a state or region of the state is 1% or less. When the percentage of infected herds are 0.2% or less than 0.1% testing may be conducted every three or 4 years respectively. In practise, most places in the UK test every 4 years. Increasing test frequency would increase the annual cost of testing and could have trade implications (Krebs et al 1997). We have explored testing intervals of 4, 3, 2, and 1 years. In an effort to explore whether more frequent testing could potentially eradicate TB in cattle we have further explored testing intervals of 6 and 8 months. All cattle moving from one farm to another are tested prior to movement and the ones detected infected are removed and slaughtered (Animal Health and Veterinary Laboratory Agencies 2013). The skin test on cattle is imperfect i.e. it's accuracy is lower than 100% and thus the test will not always detect an infected cattle. According to Defra (2009) 'studies evaluating the sensitivity of the test suggest that its sensitivity lies between 52% and 100% with median values of 80% and 93.5% for standard and severe interpretation, respectively'. Further the presence of a common parasite *Fasciola hepatica* is reported to under-ascertain the rate of the skin test to about one-third (Claridge et al. 2012). We have thus explored parameter space of skin testing accuracy of 80%, 70%, 60%, and 50% each time the test is applied. We have not included a live test for badgers as 'live test treatment was not significantly different from that in the no live test operations' (Table 5.2 in Krebs et al. 1997).

*Infections*

*Badger to cattle*
*Rationale & Parameter space*
The main route of TB infection from badgers to cattle is through inhaling or ingesting bacteria excreted by badgers directly into pasture (Benham and Broom 1991; Krebs et al. 1997). In a study using data from Woodchester park, where infections are at the upper end recorded in the UK, it was reported that badger to cattle infection rate had to be 3.4% to 6% per year in order to sustain current TB levels (Smith et al. 2001). However mean TB badger infection levels in the Smith et al. (2001) study were 16% a high value in comparison to the mean % of infected badgers in the UK which is 4.05% (Krebs et al. 1997). We have thus explored infection rates of 3.4%, 6%, and the median value 4.7% badger to cattle per infected badger individual per year. In the above, infection rates were derived by dividing the number of herd breakdowns attributed to badgers by the number of years (Smith et al. 2001) but it is not specified whether 'herd breakdowns attributed to badgers' accounted for cattle to cattle infection rate - cattle are known to infect other cattle even in the absence of badgers (Goodchild and Clifton-Hadley 2001). In other words it is not known whether in the above calculations the number of cattle infected by other cattle, if any, was removed from the analysis or whether all infections were attributed to badgers. We have therefore decided additionally to explore in simulations a badger to cattle infection rate that is a level of magnitude lower which was arbitrarily set to 0.1% per infected badger per year.

*Badger to badger*

*Model description*

Badgers infect other badgers that do not belong to the same group with a between groups badger to badger infection rate (Smith et al., 2001). The implementation is stochastic: For every infected badger on the current cell let the infection rate be a number $I_{b2b}$ and $rnd$ a uniform random number drawn in [0, 1]. If $rnd \leq I_{b2b}$ the infected badger will infect another badger on current cell. If a badger on the current cell encounters badgers from the same set it infects them with the within group badger to badger infection rate, while if not then it is infecting them with the between groups badger to badger infection rate. The process is repeated during every time step for every cell for every infected badger. The time of infection is recorded for every badger. The mean life expectancy of newly infected badgers is adjusted from mean healthy life expectancy to mean infected badger life expectancy.

*Rationale & Parameter space*

Badgers are known to be infecting each other through respiratory tract, through bites and wounds, sharing setts, as well as from mothers to cubs (Krebs et al. 1997; Jenkins et al. 2012; Smith et al. 2012). Despite the fact that there are several studies reporting the fact that badgers are infecting other badgers, there are relatively scarce data on the mechanisms of transmission (Krebs et al. 1997). In addition unless a clear mechanistic experiment is conducted where other agents of infection are isolated, it is hard to quantify the rate of infection between badgers. In a mathematical analysis of data from Woodchester Park, an area with one of the highest badger population densities as well as infection rates in the UK (Table 3.3 & 3.4 in Krebs et al. 1997) it was reported that, in the absence of other infective agents, badger to badger infection rates varied from 0.1% to 5% per year, while an infection rate of lower than 0.1% per year resulted in a failure of the disease to establish due to boundary conditions (Smith and Cheeseman 2002). In a study using sensitivity analysis of data from the same area it was concluded that within the group infection rate was 5%, but 20% between groups (Shirley et al. 2003). Given that badgers spend more than 95% of their time inside their own group territories (Roper and Lüps 1993), and that even four years after birth ≈80% of badgers were still in the groups that they are born (Woodroffe et al. 1995), and that badgers sleep together in the same chamber (Roper and Christian 1992) transmission seems most likely to occur within the sett (Krebs et al. 1997). Thus, it is unlikely that between groups infection rates will be as high as 20% but essentially this parameter is not well known. We have thus explored parameter space of within group badger to badger transmission of 0.1%, 5%, and the median value of 2.55% for every infected badger individual per year. We have further explored a level of magnitude lower than the lowest value, 0.01% to test whether such parameter space would lead to TB eradication in the absence of other infecting agents (Smith and Cheeseman 2002). We have used the same values of between group badger to badger infection rate of 0.01%, 0.1%, 2.55%, 5% (no distinction between within group and between group infections) as well as the between group values increased by 0.5% per value.

*Cattle to badger*
*Model description*

For every infected cow on current cell let the infection rate be a number $I_{c2b}$ and *rnd* a uniform random number drawn in [0, 1]. If *rnd* $\leq I_{c2b}$ the infected cow will infect a badger on current cell. The process is repeated during every time step for every cell for every infected cattle. The time of infection is recorded for every badger. The mean life expectancy of newly infected badgers is adjusted from mean healthy life expectancy to mean infected badger life expectancy.

*Rationale & Parameter space*

While there is a strong inference that cattle are infecting badgers both in terms of spatial association of *Mycabacterium bovis* in cattle and badgers (Woodroffe et al. 2005; Woodroffe et al. 2006) as well as in terms of genome sequencing (Biek et al. 2012), to our knowledge there are no quantified data on the infection rates between cattle to badgers perhaps due to the fact that the main animal of interest are cattle. However, in order to account for the full cycle of infections between and within badgers and cattle, it is important to include the four-way pattern of infections (badger to badger, badger to cattle, cattle to badger, and cattle to cattle - see also Annex 3 in King et al. 2007). This is due to the fact that badgers could be infecting cattle at a lower rate than they have been infected by cattle or vice versa and management actions should incorporate such information. We have explored a parameter space of cattle to badger infection rates of 0.1%, 3.4%, 4.7%, and 6% per infected cattle per year. In the absence of available data we used values equal to the infection rates of badger to cattle. Infection rates are essentially depended upon the total number of animals in a herd or population density within a unit of space (Goodchild and Clifton-Hadley 2001). As the number of cattle within a km² are at least 10-fold larger than the number of badgers within a km², we have used equal infection rates of cattle to badger as badger to cattle but this assumption should be rather viewed as a lower limit of the cattle to badger infection rates (i.e. cattle to badger infection rates are likely to be higher in reality assuming that the badger to cattle infection rates are the ones given).

*Cattle to cattle*

*Model description*

For every infected cow on the current cell let the infection rate be a number $I_{c2c}$ and *rnd* a uniform random number drawn in [0, 1]. If *rnd* $\leq I_{c2c}$ the infected cow will infect another cow on current cell. The process is repeated during every time step for every cell for every infected cow. The time of infection is recorded for every cow. The mean life expectancy of newly infected cattle is adjusted form mean healthy life expectancy to mean infected cattle life expectancy.

*Rationale & Parameter space*

While it is often acknowledged that TB infects both domestic and wild animals, and that wildlife can have an important role in the spreading of *Mycobacterium bovis* into cattle (Walter et al. 2012), 'some of the initial infection, and eventually most of the spread within an infected herd, may be due to cattle to cattle transmission' (Goodchild and Clifton-Hadley 2001). In the absence of infected wildlife, and badgers in particular, TB can enter the herd mainly via purchase of infected cattle (Goodchild and Clifton-Hadley 2001; Gilbert et al. 2005). Studies of TB spread within cattle in the absence of wildlife infecting agents concluded that herd size is the key factor (Denny and Wilesmith 1983; Goodchild and Clifton-Hadley 2001). This is due to the fact that the degree that cattle are infectious depends on the number of organisms excreted and the length and closeness of contact - essentially both factors depend on the number of cattle per unit of space (Costello et al. 1998). We have used cattle to cattle infection rate of 0.0073 per infected cattle per day (Barlow et al. 1997; Table 8 in Goodchild and Clifton-Hadley 2001). This value corresponds to an annual infection rate of 0.0073 x 365 = 2.6645 $\approx$ 2.7 % per infected cattle per year similar to the one given by (Conlan et al., 2012).

*Initialisation*
*Rationale & Parameter space*
Farm and herd sizes have been reported to be increasing in the UK over time (Goodchild and Clifton-Hadley 2001). According to the Royal Society for the Prevention of Cruelty to Animals (RSPCA), the average herd sizes have arisen from 71 cattle in 1994 to 92 animals in 2004 (http://www.rspca.org.uk/allaboutanimals/farm/cattle/dairy/farming). We have used a value of mean herd size of 91 animals as reported by Eurostat (2009). According to Eurostat (2009) farms in the UK vary from < 10 ha to > 200 ha per farm. We sought to quantify the effect of farm and herd size on the dynamics of TB between and within cattle by performing simulation scenarios that included mean farm size of 1, 2, 3, and 4 cells corresponding to 0.7, 1.4, 2.1, and 2.8 km² per farm respectively. Note that the initial density of cattle per km² was in all cases constant and equal to 84.7 cattle cell⁻¹. The percentage of infected badgers varies greatly depending on the study area and up to a point more studies of badgers infections may be conducted at areas with high TB prevalence in the badger population or post-mortem testing for TB prevalence can be biased towards higher TB incidents as infected badgers are more likely to be killed during culling or other occasions such as road accidents than healthy badgers (see Table 2.2 together with Table 3.4 in Krebs et al. 1997). We have parameterised the model to include initial % of infected badgers of 4.05% following the 1971-1994 UK average (N=21,731 badgers); (Table 2.2 in Krebs et al. 1997), as well as 14% simulating areas with high TB prevalence between badgers (Delahay et al. 2000). If the UK average is 4.05% then the incidence of infected badgers in some areas must be lower than 4.05%. We have thus also included simulation scenarios with initial badger infections of 2%. A study, using capture–mark–recapture over 17 years (1982–1996), of the spatial prevalence of TB in badgers (N=1270 badgers) reported that TB prevalence within the population was highly localised but also highly variable (Delahay et al. 2000). Prevalence of TB within groups (at least one infected badger within the group) varied from 33.3% to 80% (Table 3.4 in Krebs et al. 1997). As our modelling surface area covers 11,468.8 km², an area close to the surface area of a county, the prevalence of TB in badger groups (at least one infected individual within each group) cannot coincide with an overall disease prevalence of 4.05% within the badger population at that scale. We have thus explored parameter space of between groups TB prevalence (percentage of cells that contain at least one infected badger) of 35%, 40%, 45%, 50%, and 55%. We have distributed initially infected badgers in an aggregated way while cattle on a random way because badgers are social animals and so infected animals are likely to be non-randomly distributed due to other infected badgers. All simulation scenarios were replicated also in the absence of badgers (initial badger population = 0) and thus following only cattle, as well as in the absence of cattle (initial cattle population = 0) and thus following only badgers in order to quantify dynamics of TB in the absence of other infecting agents.

*Additional statistical analysis (not in main text)*

Linear mixed effects models were used to determine possible relationships between the number of infected cattle (dependent variable), and various independent variables: winter housing and the use of badger culling (as factors); and mean farm size, number of culling blocks, cattle testing interval, cattle testing accuracy, infection rate badger to badger within group, infection rate badger to badger between groups, infection rate cattle to cattle, infection rate badger to cattle, infection rate cattle to badger, initial percentage of infected cattle, mean annual cattle distance movement, mean percentage of cattle moving every year, culling intensity per cell in culling areas, percentage of cells that initially contain at least one infected badger, life span of infected badgers, badger birth rate, and initial percentage of infected badgers. The total number of cattle was used as a random effect in the model in order to account for potential larger numbers of infected cattle in large sized herds that could bias the analysis towards larger herd sizes.. Regarding mixed effects models analyses of simulation results we chose to analyse number of infected cattle as a dependent variable with total cattle as a random effect and not percentage of infected cattle within the simulated population as percentages are bounded by 0 and 100 and are sensitive to the total number of cattle individuals. We used the AIC information criterion to assess the most parsimonious model and simplify models by progressively removing the least significant covariates until no further removal was justified (Burnham and Anderson, 2002). Inspection of residual plots for constancy of variance and heteroscedasticity indicated that the models were well behaved in all cases. Analyses were conducted using the 'lme' function in R 2.14.0 using the nlme library (R Development Core Team, 2013). We implemented hierarchical variance partitioning statistical modelling of the covariates of the most parsimonious model to account for the contribution of each explanatory variable to the total variance of infected cattle (Mac Nally, 2002). Variance partitioning was conducted using the 'all.regs' and 'hier.part' functions in hier.part package in R 2.14.0 (R Development Core Team, 2013).

Mixed effects models of number of infected cattle in simulation scenarios with zero badger population (cattle only), included number of infected cattle as dependent variable, and the effects of winter housing (as factor), mean farm size, cattle testing interval, cattle testing accuracy, infection rate cattle to cattle, initial percentage of infected cattle, mean annual cattle distance movement, mean percentage of cattle moving every year. The random effect model structure included the total number of cattle.

Mixed effects models of number of infected badgers in simulation scenarios with zero cattle population (badgers only), included number of infected badgers as dependent variable, and the effects of culling (as factor), badger birth rate, badger life span when infected, culling efficiency %, number of culling blocks, and infection rate badger to badger. The random effect model structure included the initial % of infected badgers per cell nested in the initial % of infected badgers nested in the initial total number of badgers (random=~1|badgerstot/iniinfbadger/badgermaxinfpercell). The random effects structure included the three-way nesting as in the absence of cattle, the

initial % of infected badgers and the initial % of infected badgers per cell were highly correlated with the total number of badgers in each time step.

*Confronting model outputs with data*

We have identified three potential patterns to perform pattern oriented modelling: (i) Effects of culling over time in the number of infected badgers within culling areas: According to DEFRA (2010) § 3, 8, 11, 36 culling was applied for five years in an area of 100 km² with culling efficacy 70% and resulted in 28.3% decrease in TB cattle incidence at the end of that period. In the same document § 5 it was stated that at least 40-50% of cattle herd breakdowns were due to badgers in high incidence areas. We run a simulation scenario with culling applied into a block of 12 x 12 cells (12 x 12 = 144 cells x 0.7 km² per cell = 100.8 km²), with badger-to-cattle equal to cattle-to-cattle infection rates (=2.7% as recorded by Conlan et al. 2012 for cattle-to-cattle infection rates). These infection rates were set as equal in order to reproduce the fact that at least 50% of cattle herd breakdowns were due to badgers in high incidence areas. We compared the model output with the numbers regarding infected cattle after five y provided in Donnelly et al. (2011) –a more recent update of the same data as the one in DEFRA (2010). (ii) Changes in the population of badgers over time. We compared the model output in terms of initial population of badgers (65536 in all cases) vs. the final population of badgers across all scenarios - population change was calculated as [(final-initial)/initial]. We have compared the % in population change in the model against data at four published locations: Woodchester, Wytham, Bristol (Table 3.3 in Krebs et al 1997 and references therein), and the mean of several areas across the UK (Wilson et al. 1997). (iii) A scenario where cattle testing and movement was interrupted during the foot-and-mouth epidemic (FMD). Initially the model is run with parameter space as in line 9 in Supp. 2. After 20 y (month 240) of model run the model's output in terms of total and infected cattle and badgers was used as input to a simulation scenario in which all else remain the same as in the previous input but cattle testing and movement was interrupted as occurred during the 2002 FMD for 11 months.  After 11 months, the model's output in terms of total and infected cattle and badgers was used as input to a simulation scenario in which all else been equal as initially started (line 9 in Supp. 2). Model's output was compared with data from DEFRA regarding TB incidents in cattle in the UK (Incidence of TB in cattle in Great Britain – GB dataset, available at: https://www.gov.uk/government/publications/incidence-of-tuberculosis-tb-in-cattle-in-great-britain.

**Results**

The most parsimonious model included: culling as a covariate with the effect of culling to reduce the number of infected cattle (coefficient +12.25); cattle testing interval with less frequent testing associated with higher numbers of cattle infections (coefficient +193.049); mean annual cattle distance movement, with  longer distances associated with higher infections (coefficient +22.564); mean percentage of cattle moving every year with higher numbers of cattle moving resulting in higher infections (coefficient +769.484); percentage of initially infected badgers with higher initial TB prevalence in badgers resulting in higher cattle infections (coefficient +456.271); badger to cattle infection rate with higher infection rates associated with more cattle infections (coefficient

+773.311); and cattle to badger infection rate with higher infection rates associated with more cattle infections (coefficient +335.547). Cattle testing accuracy was also marginally significant but its removal was not justified and thus is included as a covariate in the final model; Higher test accuracy was associated with lower cattle infections (coefficient -56.162). These results are summarised in Table 1. All the other covariates included in the initial maximal model were removed as non-significant. Results regarding variance partitioning are listed in the main text.

The most parsimonious model of simulation scenarios with cattle only (no badgers) included: the effects of winter housing and application of winter housing increased infections (coefficient +2336.302); mean farm size with larger farms associated with more infections (coefficient +788.967); cattle testing interval with less frequent testing resulting in more infections (coefficient +100.845); mean annual cattle distance movement, with longer distances associated with higher infections (coefficient +27.544); and mean percentage of cattle moving every year with higher numbers of cattle moving resulting in higher infections (coefficient +462.764). As winter housing was implemented in the model (all cattle summoned in one cell during winter months) this was potentially likely to interact with mean farm size, an interaction effect between mean farm size and winter housing was included in the initial maximal model but dropped as non significant. These results are summarised in Table 2.

The most parsimonious model of simulation scenarios with badgers only (no cattle) included: the effects of culling efficiency and more efficient culling decreased infections (coefficient -2.364), badger life span when infected with badgers surviving longer when infected associated with more infections (coefficient +2481.055), infection rate badger to badger within group with higher within-group infection rates associated with more infections (coefficient +1452.892) and infection rate badger to badger between groups with higher between groups infections associated with more infections (coefficient +1872.663). These results are summarised in Table 3.

**Table S1.** ANOVA results of the most parsimonious model of the number of infected cattle (dependent variable). The total number of cattle was included as a random effect

|  | numDF | denDF | F-value | p-value |
|---|---|---|---|---|
| **(Intercept)** | 1 | 156 | 467.1834 | <.0001 |
| **culling** | 2 | 156 | 8.9141 | 0.0033 |
| **cowtestinterv** | 4 | 156 | 59.7986 | <.0001 |
| **cowtestaccur** | 3 | 156 | 3.8087 | 0.0528* |
| **Infrateb2cow** | 3 | 156 | 8.5877 | 0.0039 |
| **Infratecow2b** | 3 | 156 | 4.4599 | 0.0363 |
| **cowdispmean** | 4 | 156 | 54.6294 | <.0001 |
| **cowdistmean** | 4 | 156 | 7.7384 | 0.0061 |
| **Iniinfbadger** | 2 | 156 | 16.0794 | 0.0001 |

**Table S2.** ANOVA results of the most parsimonious model of the number of infected cattle (dependent variable) in simulation scenarios that included cattle only (no badgers). The total number of cattle was included as a random effect.

|  | numDF | denDF | F-value | p-value |
|---|---|---|---|---|
| **(Intercept)** | 1 | 153 | 242.8247 | <.0001 |
| **housing** | 1 | 153 | 36.26084 | <.0001 |
| **meanfarmsize** | 2 | 153 | 5.46896 | 0.0206 |
| **cowtestinterv** | 4 | 153 | 36.72815 | <.0001 |
| **cowdispmean** | 4 | 153 | 82.72493 | <.0001 |
| **cowdistmean** | 4 | 153 | 26.14834 | <.0001 |

**Table S3.** ANOVA results of the most parsimonious model of the number of infected badgers (dependent variable) in simulation scenarios that included badgers only (no cattle). The random effect model structure included the initial % of infected badgers per cell nested in the initial % of infected badgers nested in the initial total number of badgers.

|  | numDF | denDF | F-value | p-value |
|---|---|---|---|---|
| **(Intercept)** | 1 | 153 | 1302.3312 | 0.0022 |
| **badgerlifespaninf** | 2 | 153 | 195.8472 | <.0001 |
| **badgermaxcullingpercell** | 4 | 153 | 5.46896 | 0.017 |
| **infrateb2b** | 3 | 153 | 316.2157 | <.0001 |